\documentstyle[12pt]{article}
\newcommand{\be}{\begin{equation}}
\newcommand{\ee}{\end{equation}}
\begin{document}
\setlength{\unitlength}{1mm}
\vspace*{2cm}
\begin{center}
{\Large\bf Contact Interactions of Free Anyons}
\end{center}
\begin{center}
{\sc S. Voropaev }
\end{center}
\begin{center}
{\sc Vernadsky Institute,
Lab.Theoretical and Mathematical Physics, Russia Academy of Science,
Kosygina St. - 19, Moscow
}
\end{center}
\vspace*{3cm}

	"Free" ( non-scattered ) Schroedinger anyons  are considered. It
is shown that for this special type of contact interactions scattering does
not arise. By formulating the problem  for the two particles system with a
finite region of interaction  which is then allowed to go to zero, it is
established that this type of the above interactions is analog of the anomalous
magnetic moment for fermions in the Aharonov-Bohm field. Finally, we consider
connections with the Dirac anyon physics and Chern-Simons soliton solutions.

\vspace*{3cm}
cond-mat/9501015
\vspace*{3cm}

---------------------------------------------------------
\vspace*{0.1cm}

Electronic address: voropaev@tph100.physik.uni-leipzig.de

\newpage

	We suppose that the main ideas of the anyons as  quasi-particles
with  fractional spin in (2+1) space-time are known to the readers.
In other case, let us recommend $ \cite{FW}$ and $ \cite{AL}$ as  general
introduction with  exhaustive bibliography. Contact interactions of anyons
as  point particles are considered in the works  $ \cite{MT1}$ and
$ \cite{MT2}$ in particular.

\vspace*{0.1cm}

	Really, we need for our consideration  two facts only. Firstly,
it is possible to describe anyon as a particle in an Aharonov-Bohm ( AB )
gauge field,
which leads to the phase characterizing fractional statistics. Namely, we can
analyze hamiltonian and wave function $ ( h = c = 1 ) $ of the "free" point
particle $ \cite{MT1}$  $ ( V(\rho) = 0, \rho > 0 ) $ either as
\be
H = - \frac{1}{2M} \left[ \frac{\partial^{2}}{\partial \rho^{2}} +
\frac{1}{\rho} \frac{\partial}{\partial \rho} + \frac{1}{\rho^{2}}
\frac{\partial^{2}}{\partial \phi^{2}}  \right] ;
\Psi_{\theta} ( \rho, \phi + 2 \pi ) = e^{i \theta} \Psi_{\theta} ( \rho,
\phi )
\ee
or as
\be
H_{\theta} = - \frac{1}{2 M} \left[ \frac{\partial^{2}}{\partial \rho^{2}}
+ \frac{1}{\rho} \frac{\partial}{\partial \rho} + \frac{1}{\rho^{2}}
(\frac{\partial}{\partial \phi} - i \frac{\theta}{2 \pi})^{2} \right];
\Psi (\rho , \phi + 2 \pi ) = \Psi (\rho, \phi)
\ee
by means of a singular gauge transformation. Since the second  hamiltonian is
a separable operator, we can decompose the wave function $ \Psi (\vec{\rho}) $
as usual
\be
\Psi(\vec{\rho}) = \sum_{m = - \infty}^{+ \infty} C_{m} e^{i m \phi} R_{m}
(\rho)
\ee
so that the Schr$\ddot{o}$dinger equation becomes
\be
- \frac{1}{2M}  \left[ \frac{d^{2}}{d \rho^{2}} + \frac{1}{\rho}
\frac{d}{d \rho} - \frac{(\nu_{m})^{2}}{\rho^{2}}
 \right] R_{m} (\rho) = E R_{m} (\rho), \rho > 0
\ee
with
\be
\nu_{m} = \mid m - \frac{\theta}{2 \pi} \mid ; \nu = \theta / 2 \pi;
0 \leq \nu < 1
\ee
For two identical anyons $ (\rho, \phi)$ are the relative coordinates
 ( in c.c.m. ) and a rotation of $ \pi $ already leads to a physically
indistinguishable state. So, we conclude that
\be
\Psi_{\theta} ( \rho, \phi + \pi ) = e^{i \theta} \Psi_{\theta} (\rho,\phi)
\ee
and for radial wave function
\be
\nu_{m} = \mid 2m - \frac{\theta}{\pi} \mid ; \nu = \theta / \pi ;
0 \leq \nu < 1
\ee
analogous to the above consideration.

\vspace*{0.1cm}

	Secondly, in 2+1 - dimensions the equation for a "free" point Dirac
anyon $ \cite{MT2}$ is
\be
( i \hat{\partial} - e \hat{A} - M ) \Psi = 0
\ee
where $ \gamma^{0} = \sigma^{1}, \gamma^{1} = i \sigma^{2}, \gamma^{2} =
- i \sigma^{1}  $ with an  Aharonov-Bohm type magnetic
field
\be
A^{i} = - \frac{F}{2 \pi} \varepsilon^{i j} \frac{\rho^{j}}{\rho} ;
B = \varepsilon^{i j} \partial_{i} A^{j} = F \delta( \vec{\rho} )
\ee
The iteration of the Dirac equation provides
\be
[ ( - i \vec{\nabla} - e \vec{A} )^{2} - e B ] \Psi^{1}
= ( E^{2} - M^{2} ) \Psi^{1} ;
[ ( - i \vec{\nabla} - e \vec{A} )^{2} + e B ) \Psi^{2}
= ( E^{2} - M^{2} ) \Psi^{2}
\ee
These components are very similar to the above wave function with the
natural relativistic generalization : $ E^{2} \rightarrow E^{2} - M^{2} $.
But, this case of the minimal interaction is only a  particular variant
$ \theta = \pi / 2 $
of all the
possible self-adjoint extension of Dirac hamiltonian $ \cite{GR}$
general parametrization $ ( 0 \leq \theta \leq \pi; e F > 0 ) $. For
arbitrary $ \theta $, contact interacting Dirac anyons satisfy coupled
Schr$\ddot{o}$dinger equations in the interaction region. Finally, let us note
that
the correct consideration of the above problem requires a regulated contact
interaction or extended quasi-particles $ \cite{MT2} $.

\vspace*{0.1cm}

	The main aim of this paper is the analysis of the possible contact
interactions which provide a  physical interesting behavior of anyons.
One of them is the absence of  self-scattering for one anyon and
non-scattering on each other for two anyons, i.e. "soliton"-type behavior.
Let us suppose that the interaction region  is  finite  with
radius R = Const in spite of the above consideration. We investigate here
 non-relativistic particles  for simplicity, but generalization for
Dirac anyons is direct because of the remark that the components $ \Psi^{1},
\Psi^{2} $ are decoupled outside the region of interaction.

\vspace*{0.1cm}

	So, the possible wave function of anyon (5),(6) is
\be
\Psi^{(>)} (\vec{\rho}) = A \sum_{m=-\infty}^{+\infty} e^{i m \triangle \phi}
(-1)^{m} \left[ e^{ - i \pi \nu_{m} /2} J_{\nu_{m}} (p \rho) +
e^{+i \pi \nu_{m} / 2 } J_{-\nu_{m}}  (p \rho)  \right], \rho > R
\ee
where
\be
\triangle \phi \equiv \phi - \phi_{\perp} ; (\vec{p} \vec{\rho}) = p \rho
\cos( \phi - \phi_{\perp})
\ee
It is easy to show by means of the integral representation $ \cite{Sk}$
\be
\sum_{m=1}^{\infty} e^{i m \theta} J_{m+\nu}(z) = \frac{1}{2} e^{i z
\sin(\theta)} \int_{0}^{z} dx  e^{- i x \sin(\theta)} \left[ e^{i \theta}
J_{\nu} (x) + J_{\nu +1}(x) \right]
\ee
that  $ \Psi^{(>)}(\vec{\rho}) $ is
\be
\Psi^{(>)} ( \vec{\rho})= A \sin(\pi \nu) \frac{1}{2}
e^{i (\vec{p} \vec{\rho})} \int_{0}^{\infty} dx e^{- i x \cos(\triangle \phi)}
\left[ e^{+ i \pi (1 - \nu ) /2} H_{1-\nu}^{(1)} (x) - e^{+i \triangle \phi}
e^{+ i \pi \nu /2} H_{\nu}^{(1)} (x) \right]
\ee
Using the following representation $ \cite{Pr}$
\be
\int_{0}^{\infty} dx e^{- i x y} e^{+ i \pi \nu /2} H_{\nu}^{(1)} (x) =
\frac{2}{( 1 - y^{2})^{1/2} \sin(\pi \nu)} \sin
[ \nu (\pi /2 + \arcsin y ) ], \mid \nu \mid < 1, \mid y \mid < 1
\ee
we obtain the wave function of anyon
\be
\Psi^{(>)} (\vec{\rho}) = - A e^{i \nu ( \pi - \phi_{\perp})} e^{i \nu \phi}
e^{+ i ( \vec{p} \vec{\rho} )} \frac{\sin(\triangle \phi)}{\mid \sin( \triangle
\phi) \mid}
\ee
for the angles region $ \triangle \phi  \neq 0,\pi $ $ (\mid
\cos(\triangle \phi ) \mid < 1 )$ and $ 0 \leq \triangle \phi < 2 \pi $.

\vspace*{0.1cm}

	The analysis of the wave function for angles $ \triangle \phi = 0, \pi
$ is a more complex one. We mention shortly that  expressions of the type
\be
\int_{0}^{\infty} J_{\nu}(x) \left[ \begin{array}{c} \cos x  \\ \sin x
\end{array} \right] dx = ( \infty \mbox{or} 0)
\ee
are ill defined $ \cite{Pr}$ . It prompts the idea that the wave function
in this angle region is a generalized function analogous to the Dirac's
$\delta$- function. So, it is necessary to introduce  some "classical"
core for  anyon. Finally, let us note that for the
particular value $\nu = 1/2 $
("semion") and $ \triangle \phi = 0 $ : $ \Psi^{(>)} (\vec{\rho}) = 0 $,
i.e. complete reflection occurs at the origin.

\vspace*{0.1cm}

	Analogous, the wave function of the two-anyons system ( in c.c.m. )
(9) may be represented as
\be
\Psi^{(>)} (\vec{\rho}) = A  \sum_{m=1}^{\infty} e^{ + i 2 m \triangle
\phi} (-1)^{m} \left[ J_{\nu_{m}} ( p \rho) + B J_{- \nu_{m}} (p \rho )
\right] +
\ee
\be
+ A \sum_{m=0}^{\infty} e^{- i 2 m \triangle \phi} (-1)^{m} \left[ B
J_{\nu_{m}^{\prime}}(p \rho) + J_{- \nu_{m}^{\prime}} (p \rho) \right] ;
\ee
where $ \rho > R , \nu^{\prime}_{m} = 2 m + \theta / \pi, A (B) = Const $.
This expression provides quite natural wave functions for the free
two-particles
system in
(a) the bosonic limit $ ( \nu = \theta / \pi = 0 ) $:
\be
\Psi^{(>)}_{b} (\vec{\rho}) = \left[ \frac{1}{2} A (1 + B) \right]
\left[ e^{+i (\vec{p} \vec{\rho})} + e^{-i (\vec{p} \vec{\rho})} \right];
\ee
and (b) the fermionic limit $ ( \nu = \theta / \pi = 1 )$ :
\be
\Psi^{(>)}_{f} ( \vec{\rho}) = e^{+i \nu \phi} \left[ \frac{1}{2 i} A (B-1)
e^{-i \nu \phi_{\perp}} \right] \left[ e^{+i (\vec{p} \vec{\rho})} -
e^{ - i (\vec{p} \vec{\rho})} \right]
\ee
On the other hand, $ \Psi^{(>)} (\vec{\rho}) $ (18) may be transformed by
means of the integral representation  $\cite{Pr}$
\be
\sum_{k=0}^{\infty} (t)^{2 k} J_{2k + \nu} (z) = (z)^{- \nu} \int_{0}^{z}
dx x^{\nu} J_{\nu - 1} (x) \cosh \left( \frac{t x^{2}}{2 z} - \frac{tz}{2}
\right)
\ee
to the more appropriate form
\be
\Psi^{(>)} (\vec{\rho}) =e^{ +i \nu \phi} \left[ \frac{1}{2} e^{-i \nu
\phi_{\perp}} A \right] ( \left[ e^{-i \pi \nu /2} + B
e^{+i \pi \nu /2} \right] e^{-i (\vec{p} \vec{\rho})} +
\ee
\be
 + \left[ e^{+i \pi \nu /2} + B e^{-i \pi \nu /2 } \right]
e^{+i (\vec{p} \vec{\rho} )} ); 0 < \phi < \pi ; \rho > R
\ee
which coincides with the above boson and fermion wave functions for the
corresponding limit cases.

\vspace*{0.1cm}

	Using the well-known "hard-core" requirement for the anyon
interaction
 $ \cite{AL} $ , we suppose that the "back-scattering" is equal to the
"forward-scattering". So, we can conclude from the expression (23) that the
wave function of the anyon before "scattering" is
\be
\Psi^{(>)} (\vec{\rho}) = C e^{-i [ (\vec{p} \vec{\rho}) + ( E t + \triangle
/ 2 ) ] }
\ee
and after "scattering" it will be
\be
\Psi^{(>) \prime} (\vec{\rho}) = C e^{ +i [ (\vec{p} \vec{\rho}) - ( E t -
\triangle / 2 ) ] }
\ee
Here we used the condition  B $ \in {\bf R } $
for the "soliton"-type
contact interaction. ( Let us note that this condition provides self-
adjointness of the initial hamiltonian also ). It is easy to show that the
phase shift $ \triangle $ which appears after interaction of anyons with
each other is determined by $ \nu $ and  B
\be
tg (\triangle / 2) = tg (\pi \nu / 2 ) \frac{1 - B}{1 + B}
\ee
The physical interpretation of this result may be realised as a combined
contact interaction of the two anyons which includes a statistic term,
i.e. interaction of the Chern-Simons (CS) charge-flux, and term from the
CS spin-flux interaction. So, using the general expression for the phase
shift $ \triangle $ (27) we can see that the boson ( fermion ) limit
is possible now as when $ \nu \rightarrow 0 (1) $ as when $ B \rightarrow
1 (-1) $.

\vspace*{0.1cm}

	The remaining interesting problem for  applications is the
determination
of the interaction potential between two "free" anyons. It require
methods of the inverse scattering in general. We submit here a
more simple approach which is valid when $ (p R ) \rightarrow 0 $ only,
i.e in the
point quasi-particles limit. Then, we can approximate the above interaction
potential by means of the
Dirac's $ \delta $ - function, namely $ V_{int} \sim \lambda \delta ( \rho
/ R - 1 ) $. Using the well-known boundary condition following from the
properties of $ \delta $ - function
\be
R_{m} ( 1+0 ) = R_{m} ( 1-0 ) ; \frac{d R_{m}}{d x } ( 1+0 ) -
\frac{d R_{m}}{d x} ( 1-0 ) = \lambda R_{m} (1); x = \rho / R
\ee
it is easy to show from (18) that for the s-wave ( m = 0 )
\be
\lambda = - \nu + (p R / 2 )^{2 \nu } \frac{2}{B}
\frac{\Gamma ( 1 - \nu )}{\Gamma ( \nu )}
\ee
In continuation of the analogy with the particle in the AB field, we can
add to the initial hamiltonian (1) or (4) the following term
$ V_{int} = - \mu_{0} \frac{g}{2} \frac{F}{2 \pi} \frac{1}{R} \delta( \rho -
R ) $ which describes the particle with  anomalous magnetic moment ( AMM )
$  a_{e} = \frac{g - 2}{2}  $ $ \cite{BV} $. So, when $ \mu_{0} = e / 2 M;
e F / 2 \pi = \nu $ we have for the attractive potential $\lambda = - [ g/2 ]
\nu = - [ ( g-2 ) / 2 + 1 ] \nu $ and AMM will be determined by  B
\be
a_{e} = \frac{g-2}{2} = - ( p R / 2 )^{2 \nu } \frac{2}{B}
\frac{\Gamma( 1- \nu )}{\Gamma( 1 + \nu )}
\ee
For fermion with AMM in the $\delta $- shell AB magnetic field  B = - 1
which coincides with the main result of $\cite{BV}$.

\vspace*{0.1cm}

	It is clear now that we have considered the non-relativistic limit
of the Dirac equation with the non-minimal coupling by means of the
Pauli term type. On the other hand, in the relativistic
Abelian Chern-Simons theory with the Lagrange density
\be
L = \frac{k}{4} \epsilon^{\alpha \beta \gamma } A_{\alpha} F_{\beta \gamma}
+ (D_{\mu} \phi)^{*} (D_{\mu} \phi) - \frac{1}{k^{2}} \mid \phi \mid^{2}
(\mid \phi \mid^{2} - v^{2})^{2}
\ee
topological solitons/vortices with  quantized
flux and  non-topological solitons with no quantized flux are possible
$ \cite{J} $
\be
\nabla^{2} \ln \mid \phi \mid^{2} = - \frac{4 v^{2}}{k^{2}} \mid \phi
\mid^{2} ( 1 - \mid \phi \mid^{2} / v^{2} )
\ee
The later are especially interesting for us because of the
appropriate matter field distribution as in  our $\delta$- shell model
\be
\mid \phi \mid^{2} \rightarrow \rho^{2 N - 2}, \rho \rightarrow 0 ;
\mid \phi \mid^{2} \rightarrow \rho^{-2(\alpha + 1)}, \rho \rightarrow
\infty,
\ee
where $\alpha \geq N $, $ N = 1,2,... $ is a soliton number. These solutions
posses angular momentum $ J = k F / 2 ( \alpha - N) $ where F is CS flux.
One may consider the  motion of the non-topological solitons and derive
an effective Lagrangian from the underlaying field theory. It is reasonable
that as in topological soliton considerations $\cite{K}$ one finds a
statistical interaction term with the value of the statistical factor in
agreement with the generalized spin-statistic relation. So, it is possible
that  our consideration of the contact interaction of "free" anyons
will  provide a quantum version of the non-topological Abelian CS solitons.

\vspace*{1cm}

	I would like to thank Dr. M. Bordag, Dr. M. Hellmund
for helpful
conversations and the Institute Theoretical Physics, Leipzig University for
kind
hospitality. Also, I want to express my deep gratitude Dr. R. Herrmann
for the support and as an initiator of a number of ideas.

\end{document}